\documentclass[12pt]{article}
\usepackage{graphicx}
\begin{document}

\begin{center}
{\bf Minimal energy configurations for charged particles on a thin conducting disc: the perimeter particles

\vspace{10pt}
A. Worley}

\vspace{10pt}
{\it HH Wills Physics Laboratory, University of Bristol 

Tyndall Avenue, Bristol BS8 1TL, UK
\vspace{10pt}

(aw@xiboo.co.uk)}
\end{center}
\vspace{10pt}
\begin{abstract}
The lowest energy configurations for $N$ equal charged particles confined to a thin conducting disc have been investigated in detail up to $N=160$ and in outline for further values up to $N=500$.  For all values of $N$ up to 160 the particle configurations can be described in terms of concentric shells.  The number of perimeter particles $p$ appears to be simply related to $N$ and to the mean radius of the outermost internal shell. Justification for these relations is obtained from a simple model based on the well-known distribution of continuous charge on a conducting disc.
\end{abstract}
\newpage
\section{Introduction}
If $N$ identical charged particles are confined to the interior of a conducting hollow sphere, all the particles take up positions on the sphere itself - there is no stable, static equilibrium involving any particle with $r<1\;$[1]. The problem of how the particles arrange themselves on the sphere was originally posed by Thomson [2] and is still of great interest [see for example 3,4].  The corresponding problem for $N$ similar particles confined to a thin conducting disc has a more recent history. Given that in the sphere case all the particles must reside on the boundary, it is perhaps tempting to assume that in the 2-D problem the $N$ particles will take up equally-spaced positions around the perimeter.  For small values of $N$, this is exactly what is observed, but in 1985 Berezin [5] reported the \lq unexpected\rq \hspace{3pt} result that for the lowest energy configurations with $N>11$, the particles do not all reside on the perimeter - it is energetically favourable for some of the particles to be positioned in the interior. It was quickly pointed out, however, [6,7] that for very high $N$ the configuration of discrete charges should approach the distribution for continuous charge, for which the result is well known and does indeed feature a non-zero charge density at all radii. 

For continuous charge, the charge density  $\sigma (r)$ is calculated with great economy by Friedberg [8].  The equilibrium distribution of charge across the disc is shown to be equivalent to that found by projecting the uniform distribution on a sphere onto its equatorial plane.  The result, also derived by conventional methods [9] is
$$\sigma (r)={2\sigma (0)\over {(1-r^2)^{1/2}}}\eqno(1)$$
where $\sigma (0)$ is the central charge density.  The charge density  becomes infinite at $r=1$.

For discrete charges, at least for low values of $N$, the particles are arranged in a system of concentric shells.  Detailed calculations for $N\leq 80$ have been performed by Nurmela [10], who reviews earlier work and has reported the energies for each value of $N$ at what is thought to be its optimal configuration.  Oymak and Erko\c c [11] have also reported configurations and energies for $N\leq 90$.  There is also a body of related work [see for example 12,13] in which the particles are confined to the disc by a parabolic potential ($V(r)={1/2}m{{\omega _0}^2}{r^2}$) rather than by hard-wall confinement ($r<1,V(r)=0;\: r\ge 1,V(r)=\infty$). Here we use hard-wall confinement only and study the range ${21\leq N\leq 160}$ in detail and $170\leq N\leq 500$ in outline, obtaining a relationship between the number $p$ of perimeter particles and $N$. 

\section{Methods}
\subsection{Basic Methods}
All the programs used in this work employ the same relaxation routine.  An initial distribution of particles is generated within the unit circle and the distances between all pairs calculated.   The net Coulomb force on the $i$th particle is
$${\rm \bf F}_i=\sum_{j\neq i}^N {{\rm \bf r}_{ij} \over r_{ij}^3}\eqno(2)$$						
and each particle is then moved, subject to remaining within the perimeter, in the direction of the net force and by a distance chosen to be equal to $F/N^{\rm 2}$.  The procedure is then repeated and periodically the total energy
$$E(N)=\sum_{i<j}^N{1 \over {r_{ij}}}\eqno(3)$$											 
is calculated.  In the simplest version of the program, execution is halted when the total energy fails to improve by a fixed amount (for example, $\Delta E=0.0001$) over 1000 iterations.  As $N$ increases the calculation of the interparticle forces eventually dominates the processing, requiring a time $O(N^2)$, so in some versions of the program there are optional survey modes in which $\Delta E$ is increased to 0.001 or even 0.01.
 
The system of $N$ particles may encounter a local energy minimum from which no further progress may be made.  Fig.1 shows part of the spectrum of such metastable states for $N=100$, organised by the number of particles on the perimeter $p(N)$.  These alternative endpoints can account for a large fraction of the computing resources if no steps are taken to limit them.  The approach taken here has been first to restrict $p(N)$ to a specified range, but only after an initial survey has established the most likely $p$ value.  Some versions of the program also employ a simple form of annealing in which the positions of the interior particles are repeatedly disrupted with the aim of avoiding local energy minima.  

For programs running with a specified value of $p$, the required number of particles are set uniformly around the perimeter and the remaining $N-p$ at random in the interior.  If no particular value of $p$ is specified, however, care must be taken with the initial distribution because the perimeter count tends to freeze early in the calculation.  As Nurmela [10] points out, particles arriving first at the perimeter tend to inhibit further particles from joining them.  If the initial distribution is totally random, the final states tend to be drawn from those on the left of spectra such as Fig. 1.  If all particles are placed initially on the perimeter, the final states are biased to those on the right of the diagram.  To achieve a range of $p$ values centred close to the optimal value, we have employed initial radial distributions of the form $r=1-\lambda *{\rm random}$ where the factor $\lambda \sim 0.1$ must be adjusted slightly for $N$.

\begin{figure}
\begin{center}
\includegraphics[scale=0.45]{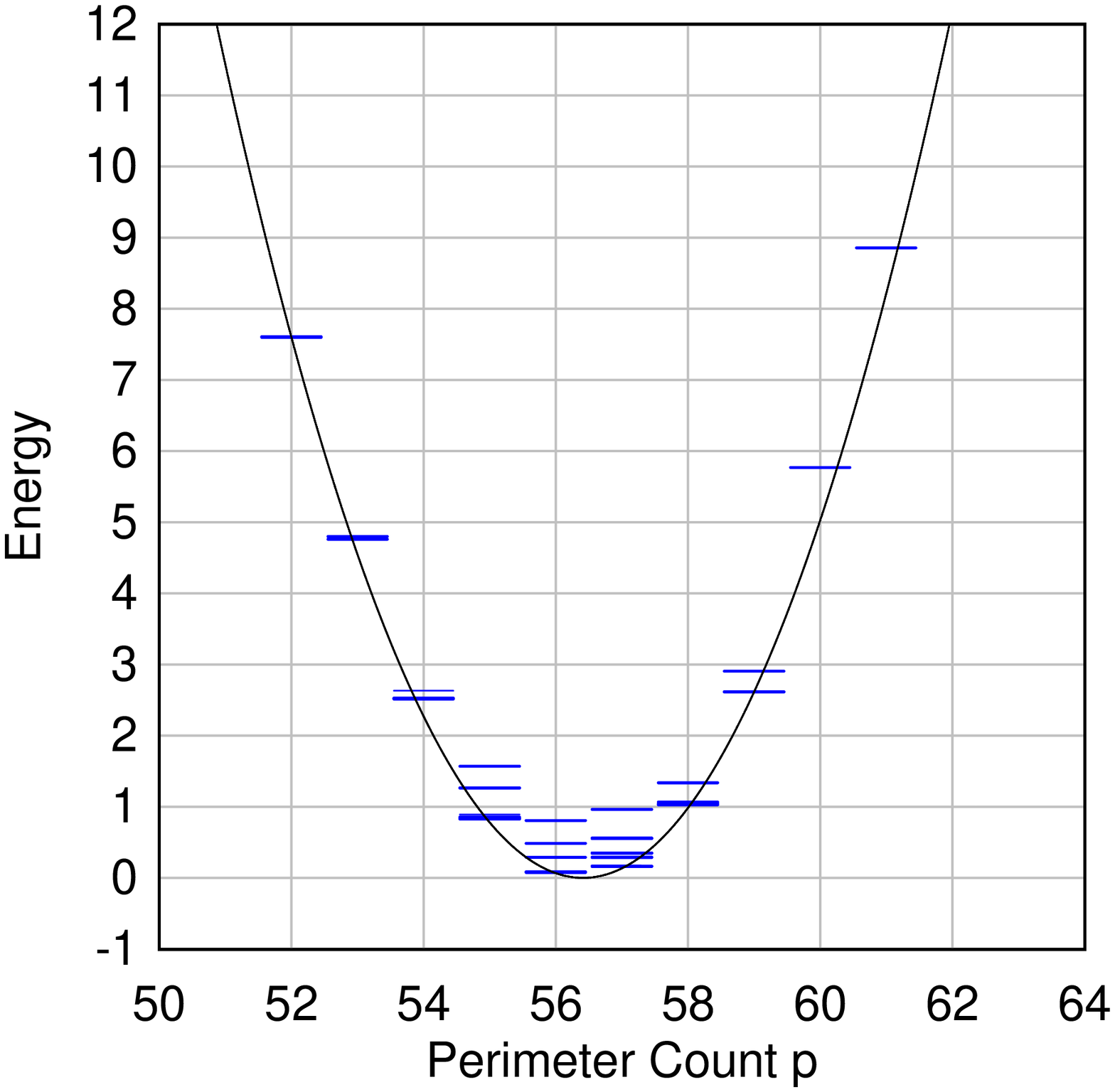}
\vspace{10pt}
\begin{minipage}[c]{0.8\linewidth}
\noindent \small Fig.1. Part of the spectrum of final states for $N=100$, organised by $p$-value.  The parabola traces the envelope of the lowest energies for each $p$ and defines the zero on the energy scale.
\end{minipage}
\end{center}
\end{figure}

\subsection{Three Stages of Computation}
The initial stage of computation was concerned only in the preparation of a list of candidate values of $p(N)$.  For ${21\leq N\leq 80}$, these were taken from the well-established results of [10].  For ${81\leq N\leq 160}$, various early versions of the program described below were used.  Support for the full list of candidates was then obtained by generating an energy spectrum of final states similar to Fig.1 and for each $N$.  Energy determinations of the highest accuracy were not sought and a basic program without preselection of $p$ and terminating at $\Delta E=0.01$ was used.  In every case the envelope of the lowest energies was parabolic in $p$ and was fitted to $E=E_{\rm min}+\alpha {(p-p_{\rm min})}^2$.  Here, $E_{\rm min}$ is a lower energy than the optimal value associated with an integer $p$-value, and plays no further part in this work. The values of $p_{\rm min}$ are included in Table 1 and the variation of $\alpha$ with $N$ is shown in Fig.2.
  
The candidate integer values of $p$ were then used in the main stage of computation.  For each $N$ in the range ${21\leq N\leq 160}$ a calculation was performed with a predetermined perimeter count set in turn to $p-1$, $p$ and $p+1$.  For each $p$-value, an initial distribution was generated which was then allowed to evolve during 100 sessions each of 10000 iterations.  Between sessions, the position of each internal particle was modified by a random vector with amplitude $s*{\rm random}$. At each annealing stage, the value of $s$ was reduced by $1\%$ of its initial value, which was taken as 0.4.  This choice meant that in the initial stages for most $N$ values, several of the internal particles moved outside the perimeter, forcing the generation of a new distribution.  This annealing procedure is clearly less sophisticated than the formal simulated annealing algorithm first applied to this problem by Wille and Vennik [14], which requires that each particle is moved in turn and the new configuration is then accepted or not on the basis of the change in energy.

\begin{figure}
\begin{center}
\includegraphics[scale=0.4]{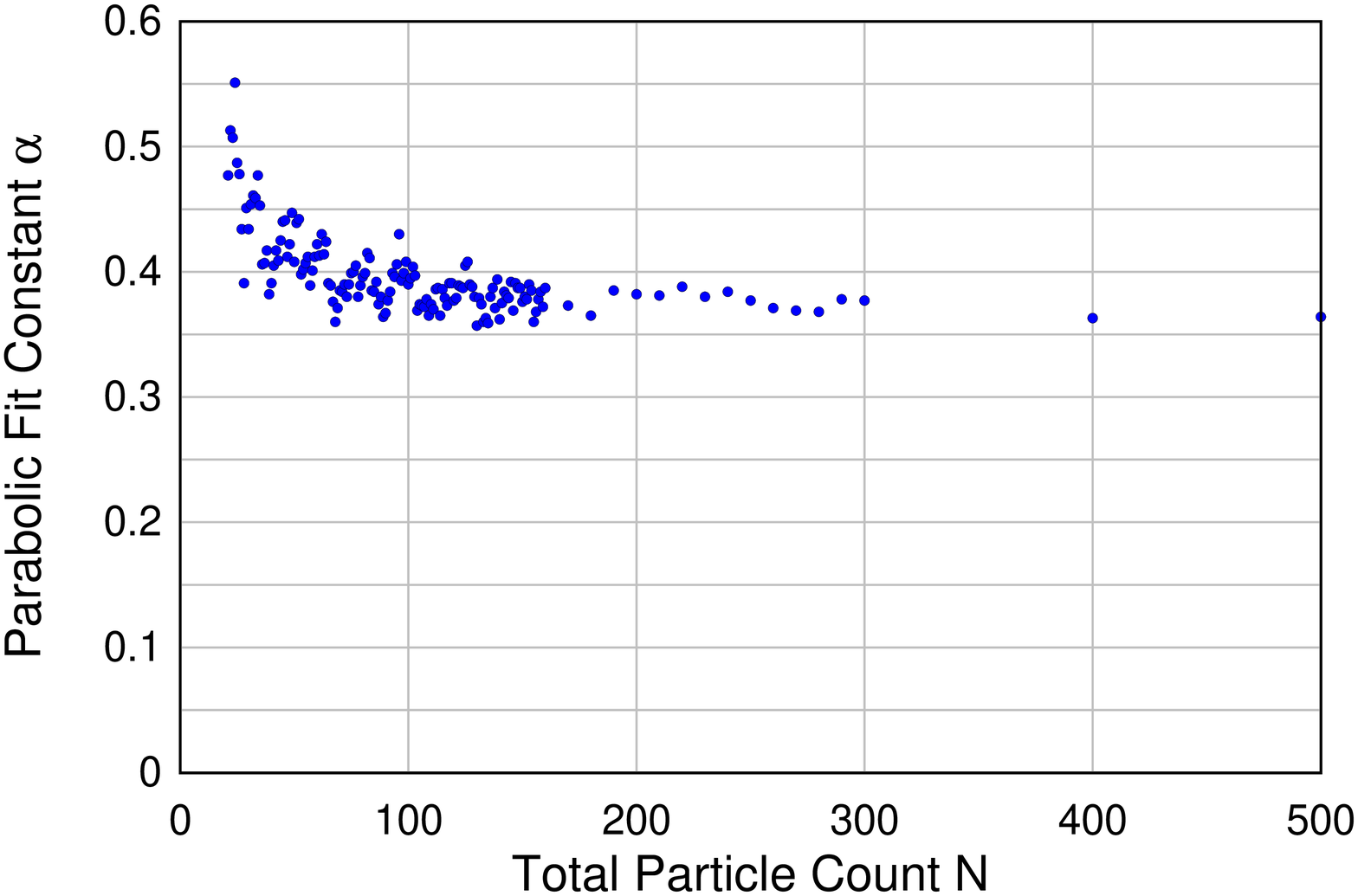}
\vspace{10pt}
\begin{minipage}[c]{0.8\linewidth}
\noindent \small Fig.2. Variation of the parabolic fit parameter $\alpha$ with $N$ for ${21\leq N\leq 160}$ and for selected values up to $N=500$.  The typical standard deviation for repeated measurements on a value of $\alpha$ is 0.01.
\end{minipage}
\end{center}
\end{figure}
  
The confirmation stage was a repeat of the main stage.  The only data carried forward from the main stage were the $p$-values of the best results, but this time the variants with $p-1$ and $p+1$ were not calculated.  The confirmation stage was performed twice, and the two sets of results were compared with the results of the previous stage.  The lowest energy of the three results for each $N$ is shown in Table 1.  A further run was then performed with an alternative form of annealing - instead of $s=0.4$, $s$ was taken as $1-r$, where $r$ was the radial distance of each internal particle from the centre. This produces a less extreme disturbance of the particle positions, but was not found to offer any advantage.  Though the bulk of the results were identical with those already obtained, 21 had marginally higher energies and only for two values of $N$ was a small improvement (one digit in the final quoted decimal place) obtained.

\subsection {Results for ${21\leq N\leq 160}$ }
For all ${21\leq N\leq 80}$ the lowest energy was produced at the central $p$-value tested, and the energies and configurations agree in every particular with those of [10].  The results in this range are listed here for completeness and to include the measurements of $p_{\rm min}$ and of the mean radius of the outermost inner shell $r_{\rm 2}$.     
In the range ${81\leq N\leq 160}$, for all but one value of $N$ the lowest energy found in the main stage was again produced by the central $p$-value of the three tested, and the energy determined was equal to or slightly below (one or two places in the final figure) the value noted at the survey stage.  The small reductions reflect the superior performance of the 100 restart/annealing stages compared with earlier versions of the program.

\subsection {Results for $N>160$}
For N=170,180...290 the spectrum program was run without further investigation, and the values of $\alpha$ and $p_{\rm min}$ only are shown in Table 1.  The parabola constant $\alpha$ (Fig.2) is well-behaved, and appears to settle to a value of around 0.36 as $N$ is increased, suggesting that the envelopes of all spectra eventually have a similar shape. This predictability is supported by the good agreement between $p_{\rm min}$ and the optimal value of $p(N)$ for $N\leq 160$ (Table 1) and suggests that only a limited range of $p$-values near $p_{\rm min}$ need be investigated if computing resources are at a premium.  Accordingly, for $N$ = 200, 300, 400 and 500 the spectrum program was first used to determine $p_{\rm min}$ and the annealing program was then run for one or two candidate values of $p$.  Full configurations were not obtained for these values and the listed energies are not necessarily optimal.

\section{Shell Structure}
Three points were noted:

\noindent(1) In the range $N\le 80$, the configuration for $N+1$ is, with one exception, the same as that for $N$ with the addition of one particle to an existing shell or the creation of a new shell with a particle placed at the centre.  The single exception is the pair $N=55$ (5-13-37) and $N=56$ (1-6-12-37).  For the range $81\leq N\leq160$ there are four exceptions to the usual pattern: $N$ = 97/98, 117/118,  150/151 and 152/153.

\noindent(2) Oymak and Erko\c c [11] have suggested that new shells are created when $N$ is given by
$$N=12+\sum_{n=1}^m(10+8n)\eqno(4)$$ 
This is equivalent to N=(2m+3)(2m+4) and suggests that new shells are created at $N=12, 30, 56, 90, 132, 182...$.  The results obtained here, however, show the next new shell after $N=90$ appearing at $N=134$.

\noindent(3) Of the five possible 2D Bravais lattices, it is the triangular lattice that produces the lowest energy packing for identical charged particles [15], and this arrangement is observed in the central region for many values of $N$. Nevertheless, for $N\leq 90$ previous authors have also described the optimal configurations in terms of concentric shells, and this property appears to be maintained up to $N=160$.  Beyond this value, the first evidence of a departure from this behaviour may occur at $N=185$ (Fig.3).  Repeated attempts have been made to improve upon this structure and its energy ($E=23129.9182$) but no lower energy state has been found.  $N=185$ may be an initial, isolated example of a later trend, because preliminary results for the remaining $N$ in the range ${161\leq N\leq 200}$ all appear to have well-ordered shell structures.

\newpage
\begin{center}
{\bf Table 1.  Results for ${\bf 21\leq N\leq 72}$}

------------------------------------------------------------------------

{\footnotesize \  $N$ \ Energy $E(N)$ \ \ \ \ \ Configuration \ \ \ \ \ \ \ $p_{\rm min}$ \ \ \ \ $\alpha$ \ \ \ \ \ \ \ $r_{\rm 2}$ \ \ \ }

------------------------------------------------------------------------

{\scriptsize{
\begin{tabular}{ccrrcccccccc}
        21 &  216.17911 &            &            &            &            &          3 &         18 &            &      17.72 &      0.477 &    0.37890 \\

        22 &  240.12168 &            &            &            &            &          4 &         18 &            &      18.40 &      0.513 &    0.43825 \\

        23 &  265.20104 &            &            &            &            &          4 &         19 &            &      18.99 &      0.507 &    0.43146 \\

        24 &  291.72783 &            &            &            &            &          4 &         20 &            &      19.73 &      0.551 &    0.42511 \\

        25 &  319.66551 &            &            &            &            &          5 &         20 &            &      20.31 &      0.487 &    0.47061 \\

        26 &  348.77089 &            &            &            &            &          5 &         21 &            &      20.86 &      0.478 &    0.46427 \\

        27 &  379.35332 &            &            &            &            &          5 &         22 &            &      21.57 &      0.434 &    0.45827 \\

        28 &  411.34431 &            &            &            &            &          6 &         22 &            &      22.28 &      0.391 &    0.49515 \\

        29 &  444.54775 &            &            &            &            &          6 &         23 &            &      22.70 &      0.451 &    0.48923 \\

        30 &  479.07957 &            &            &            &          1 &          6 &         23 &            &      23.39 &      0.434 &    0.54870 \\

        31 &  514.91713 &            &            &            &          1 &          6 &         24 &            &      23.97 &      0.454 &    0.54275 \\

        32 &  552.26974 &            &            &            &          1 &          6 &         25 &            &      24.55 &      0.461 &    0.53707 \\

        33 &  590.80630 &            &            &            &          1 &          7 &         25 &            &      25.08 &      0.459 &    0.55861 \\

        34 &  630.83438 &            &            &            &          1 &          7 &         26 &            &      25.68 &      0.477 &    0.55312 \\

        35 &  672.34218 &            &            &            &          1 &          8 &         26 &            &      26.23 &      0.453 &    0.57278 \\

        36 &  715.06832 &            &            &            &          1 &          8 &         27 &            &      26.94 &      0.406 &    0.56747 \\

        37 &  759.34275 &            &            &            &          1 &          8 &         28 &            &      27.44 &      0.407 &    0.56237 \\

        38 &  804.92000 &            &            &            &          2 &          8 &         28 &            &      27.96 &      0.417 &    0.59887 \\

        39 &  851.91130 &            &            &            &          2 &          8 &         29 &            &      28.57 &      0.382 &    0.61107 \\

        40 &  900.11872 &            &            &            &          2 &          9 &         29 &            &      29.07 &      0.391 &    0.60794 \\

        41 &  949.85158 &            &            &            &          2 &          9 &         30 &            &      29.57 &      0.405 &    0.60312 \\

        42 &  1000.8254 &            &            &            &          3 &          9 &         30 &            &      30.08 &      0.417 &    0.64265 \\

        43 &  1053.3045 &            &            &            &          3 &          9 &         31 &            &      30.57 &      0.409 &    0.62752 \\

        44 &  1107.0794 &            &            &            &          3 &         10 &         31 &            &      31.09 &      0.425 &    0.63813 \\

        45 &  1162.3304 &            &            &            &          3 &         10 &         32 &            &      31.63 &      0.440 &    0.63361 \\

        46 &  1218.9808 &            &            &            &          4 &         10 &         32 &            &      32.11 &      0.441 &    0.65754 \\

        47 &  1277.0070 &            &            &            &          4 &         10 &         33 &            &      32.67 &      0.412 &    0.65316 \\

        48 &  1336.3752 &            &            &            &          4 &         11 &         33 &            &      33.22 &      0.422 &    0.66146 \\

        49 &  1397.2009 &            &            &            &          4 &         11 &         34 &            &      33.71 &      0.447 &    0.65722 \\

        50 &  1459.5821 &            &            &            &          5 &         11 &         34 &            &      34.14 &      0.408 &    0.67731 \\

        51 &  1523.2092 &            &            &            &          5 &         11 &         35 &            &      34.73 &      0.439 &    0.67321 \\

        52 &  1588.2065 &            &            &            &          5 &         12 &         35 &            &      35.23 &      0.442 &    0.67998 \\

        53 &  1654.6578 &            &            &            &          5 &         12 &         36 &            &      35.71 &      0.398 &    0.67601 \\

        54 &  1722.6842 &            &            &            &          5 &         13 &         36 &            &      36.17 &      0.402 &    0.68265 \\

        55 &  1791.9737 &            &            &            &          5 &         13 &         37 &            &      36.62 &      0.407 &    0.67880 \\

        56 &  1862.6497 &            &            &          1 &          6 &         12 &         37 &            &      37.06 &      0.412 &    0.70499 \\

        57 &  1934.7385 &            &            &          1 &          6 &         13 &         37 &            &      37.61 &      0.389 &    0.70966 \\

        58 &  2008.0682 &            &            &          1 &          6 &         13 &         38 &            &      38.08 &      0.401 &    0.70597 \\

        59 &  2083.0334 &            &            &          1 &          6 &         13 &         39 &            &      38.56 &      0.412 &    0.70236 \\

        60 &  2159.3584 &            &            &          1 &          6 &         14 &         39 &            &      38.98 &      0.422 &    0.70720 \\

        61 &  2237.1926 &            &            &          1 &          6 &         14 &         40 &            &      39.57 &      0.413 &    0.70369 \\

        62 &  2316.2518 &            &            &          1 &          7 &         14 &         40 &            &      40.03 &      0.430 &    0.71683 \\

        63 &  2396.9504 &            &            &          1 &          7 &         14 &         41 &            &      40.49 &      0.414 &    0.71343 \\

        64 &  2478.9810 &            &            &          1 &          7 &         15 &         41 &            &      40.98 &      0.424 &    0.71780 \\

        65 &  2562.5685 &            &            &          1 &          7 &         15 &         42 &            &      41.52 &      0.391 &    0.71448 \\

        66 &  2647.4781 &            &            &          1 &          8 &         15 &         42 &            &      41.94 &      0.389 &    0.72632 \\

        67 &  2733.9499 &            &            &          1 &          8 &         15 &         43 &            &      42.44 &      0.376 &    0.72310 \\

        68 &  2821.6862 &            &            &          1 &          8 &         16 &         43 &            &      42.88 &      0.360 &    0.72674 \\

        69 &  2910.8539 &            &            &          2 &          8 &         16 &         43 &            &      43.28 &      0.371 &    0.73711 \\

        70 &  3001.4558 &            &            &          2 &          8 &         16 &         44 &            &      43.68 &      0.385 &    0.73401 \\

        71 &  3093.5796 &            &            &          2 &          9 &         16 &         44 &            &      44.11 &      0.384 &    0.74411 \\

        72 &  3187.0854 &            &            &          2 &          9 &         16 &         45 &            &      44.58 &      0.390 &    0.74110 \\
\end{tabular}}

\normalsize
 
------------------------------------------------------------------------}
\end{center}
\newpage
\begin{center}
{\bf Table 1 (cont.) Results for ${\bf 73\leq N\leq 124}$}

-----------------------------------------------------------------------------

{\footnotesize \ \ $N$ \ Energy $E(N)$ \ \ \ \ \ \ \ \ Configuration \ \ \ \ \ \ \ \ $p_{\rm min}$ \ \ \ \ $\alpha$ \ \ \ \ \ \ \ \ $r_{\rm 2}$ \ \ \ }

-----------------------------------------------------------------------------

{\scriptsize{\begin{tabular}{ccrrcccccccc}
        73 &  3281.8499 &            &            &          3 &          9 &         16 &         45 &            &      45.00 &      0.380 &    0.75073 \\

        74 &  3378.1982 &            &            &          3 &          9 &         17 &         45 &            &      45.43 &      0.390 &    0.75316 \\

        75 &  3475.8612 &            &            &          3 &          9 &         17 &         46 &            &      45.88 &      0.399 &    0.75025 \\

        76 &  3575.2077 &            &            &          3 &          9 &         17 &         47 &            &      46.31 &      0.400 &    0.74739 \\

        77 &  3675.7938 &            &            &          3 &         10 &         17 &         47 &            &      46.71 &      0.405 &    0.75599 \\

        78 &  3777.8993 &            &            &          3 &         10 &         18 &         47 &            &      47.24 &      0.380 &    0.75847 \\

        79 &  3881.4275 &            &            &          3 &         10 &         18 &         48 &            &      47.66 &      0.389 &    0.75570 \\

        80 &  3986.2335 &            &            &          4 &         10 &         18 &         48 &            &      48.09 &      0.396 &    0.76366 \\

        81 &  4092.6937 &            &            &          4 &         10 &         18 &         49 &            &      48.50 &      0.399 &    0.76097 \\

        82 &  4200.5465 &            &            &          4 &         11 &         18 &         49 &            &      48.97 &      0.415 &    0.76857 \\

        83 &  4309.8590 &            &            &          4 &         11 &         19 &         49 &            &      49.36 &      0.411 &    0.77050 \\

        84 &  4420.5247 &            &            &          4 &         11 &         19 &         50 &            &      49.84 &      0.385 &    0.76790 \\

        85 &  4532.7613 &            &            &          5 &         11 &         19 &         50 &            &      50.23 &      0.384 &    0.77496 \\

        86 &  4646.3751 &            &            &          5 &         11 &         19 &         51 &            &      50.64 &      0.392 &    0.77242 \\

        87 &  4761.4694 &            &            &          5 &         11 &         20 &         51 &            &      51.06 &      0.374 &    0.77425 \\

        88 &  4878.0585 &            &            &          5 &         11 &         20 &         52 &            &      51.44 &      0.380 &    0.77177 \\

        89 &  4995.9139 &            &            &          5 &         12 &         20 &         52 &            &      51.91 &      0.364 &    0.77832 \\

        90 &  5115.4077 &            &          1 &          5 &         12 &         20 &         52 &            &      52.27 &      0.367 &    0.78451 \\

        91 &  5236.2113 &            &          1 &          5 &         12 &         20 &         53 &            &      52.68 &      0.377 &    0.78212 \\

        92 &  5358.3535 &            &          1 &          6 &         12 &         20 &         53 &            &      53.09 &      0.384 &    0.78810 \\

        93 &  5482.1253 &            &          1 &          6 &         12 &         20 &         54 &            &      53.50 &      0.399 &    0.78576 \\

        94 &  5607.1884 &            &          1 &          6 &         12 &         21 &         54 &            &      53.89 &      0.396 &    0.78718 \\

        95 &  5733.8395 &            &          1 &          6 &         13 &         21 &         54 &            &      54.33 &      0.406 &    0.79285 \\

        96 &  5861.8665 &            &          1 &          6 &         13 &         21 &         55 &            &      54.73 &      0.430 &    0.79057 \\

        97 &  5991.4704 &            &          1 &          6 &         13 &         22 &         55 &            &      55.15 &      0.393 &    0.79192 \\

        98 &  6122.4986 &            &          1 &          7 &         13 &         21 &         56 &            &      55.55 &      0.399 &    0.79381 \\

        99 &  6254.8303 &            &          1 &          7 &         13 &         22 &         56 &            &      55.99 &      0.408 &    0.79503 \\

       100 &  6388.7591 &            &          1 &          7 &         14 &         22 &         56 &            &      56.41 &      0.390 &    0.80009 \\

       101 &  6524.0431 &            &          1 &          7 &         14 &         22 &         57 &            &      56.82 &      0.395 &    0.79793 \\

       102 &  6660.9258 &            &          1 &          7 &         14 &         23 &         57 &            &      57.25 &      0.404 &    0.79910 \\

       103 &  6799.2266 &            &          1 &          7 &         14 &         23 &         58 &            &      57.58 &      0.397 &    0.79697 \\

       104 &  6939.1183 &            &          1 &          8 &         14 &         23 &         58 &            &      58.06 &      0.369 &    0.80189 \\

       105 &  7080.4212 &            &          1 &          8 &         14 &         23 &         59 &            &      58.45 &      0.374 &    0.79981 \\

       106 &  7222.9971 &            &          2 &          8 &         14 &         23 &         59 &            &      58.79 &      0.373 &    0.80442 \\

       107 &  7367.0551 &            &          2 &          8 &         15 &         23 &         59 &            &      59.19 &      0.372 &    0.80896 \\

       108 &  7512.6343 &            &          2 &          8 &         15 &         23 &         60 &            &      59.56 &      0.378 &    0.80695 \\

       109 &  7659.5604 &            &          2 &          8 &         15 &         24 &         60 &            &      59.95 &      0.365 &    0.80787 \\

       110 &  7808.1538 &            &          3 &          8 &         15 &         24 &         60 &            &      60.31 &      0.374 &    0.81213 \\

       111 &  7958.0348 &            &          3 &          8 &         15 &         24 &         61 &            &      60.68 &      0.370 &    0.81018 \\

       112 &  8109.3152 &            &          3 &          9 &         15 &         24 &         61 &            &      61.09 &      0.386 &    0.81437 \\

       113 &  8262.2155 &            &          3 &          9 &         15 &         24 &         62 &            &      61.47 &      0.387 &    0.81245 \\

       114 &  8416.4049 &            &          3 &          9 &         16 &         24 &         62 &            &      61.83 &      0.365 &    0.81647 \\

       115 &  8572.0984 &            &          3 &          9 &         16 &         25 &         62 &            &      62.19 &      0.386 &    0.81721 \\

       116 &  8729.3215 &            &          3 &          9 &         16 &         25 &         63 &            &      62.58 &      0.379 &    0.81533 \\

       117 &  8888.1133 &            &          3 &          9 &         17 &         25 &         63 &            &      63.04 &      0.373 &    0.81918 \\

       118 &  9048.3699 &            &          3 &         10 &         16 &         25 &         64 &            &      63.36 &      0.391 &    0.81739 \\

       119 &  9209.9008 &            &          4 &         10 &         16 &         25 &         64 &            &      63.76 &      0.391 &    0.82112 \\

       120 &  9372.9075 &            &          4 &         10 &         16 &         26 &         64 &            &      64.12 &      0.377 &    0.82176 \\

       121 &  9537.4172 &            &          4 &         10 &         17 &         26 &         64 &            &      64.52 &      0.379 &    0.82536 \\

       122 &  9703.2866 &            &          4 &         10 &         17 &         26 &         65 &            &      64.90 &      0.389 &    0.82359 \\

       123 &  9870.8678 &            &          4 &         10 &         17 &         27 &         65 &            &      65.24 &      0.388 &    0.82421 \\

       124 &  10039.801 &            &          4 &         10 &         17 &         27 &         66 &            &      65.65 &      0.387 &    0.82246 \\
\end{tabular}}

\normalsize
-----------------------------------------------------------------------------}
\end{center}
\newpage
\begin{center}
{\bf Table 1 (cont.)  Results for ${\bf 125\leq N\leq 160}$ and selected values ${\bf N\leq 500}$}

------------------------------------------------------------------------------

{\footnotesize \ \ $N$ \ Energy $E(N)$ \ \ \ \ \ \ \ \ Configuration \ \ \ \ \ \ \ \ \ \ $p_{\rm min}$ \ \ \ \ \ \ $\alpha$ \ \ \ \ \ \ \ \ $r_{\rm 2}$ \ \ \ }

------------------------------------------------------------------------------

{\scriptsize{\begin{tabular}{ccrrcccccccc}
       125 &  10210.214 &            &          4 &         10 &         18 &         27 &         66 &            &      66.02 &      0.405 &    0.82588 \\

       126 &  10382.103 &            &          4 &         11 &         18 &         27 &         66 &            &      66.39 &      0.408 &    0.82921 \\

       127 &  10555.369 &            &          4 &         11 &         18 &         27 &         67 &            &      66.80 &      0.390 &    0.82750 \\

       128 &  10730.222 &            &          5 &         11 &         18 &         27 &         67 &            &      67.17 &      0.388 &    0.83079 \\

       129 &  10906.538 &            &          5 &         11 &         18 &         27 &         68 &            &      67.50 &      0.380 &    0.82912 \\

       130 &  11084.238 &            &          5 &         11 &         18 &         28 &         68 &            &      67.85 &      0.357 &    0.82966 \\

       131 &  11263.536 &            &          5 &         11 &         19 &         28 &         68 &            &      68.27 &      0.379 &    0.83275 \\

       132 &  11444.221 &            &          5 &         11 &         19 &         28 &         69 &            &      68.62 &      0.374 &    0.83112 \\

       133 &  11626.331 &            &          5 &         12 &         19 &         28 &         69 &            &      68.95 &      0.360 &    0.83415 \\

       134 &  11809.947 &          1 &          5 &         12 &         19 &         28 &         69 &            &      69.29 &      0.363 &    0.83710 \\

       135 &  11994.978 &          1 &          5 &         12 &         19 &         28 &         70 &            &      69.60 &      0.359 &    0.83550 \\

       136 &  12181.345 &          1 &          6 &         12 &         19 &         28 &         70 &            &      70.00 &      0.380 &    0.83842 \\

       137 &  12369.335 &          1 &          6 &         12 &         19 &         29 &         70 &            &      70.36 &      0.387 &    0.83884 \\

       138 &  12558.742 &          1 &          6 &         12 &         19 &         29 &         71 &            &      70.73 &      0.371 &    0.83728 \\

       139 &  12749.699 &          1 &          6 &         12 &         20 &         29 &         71 &            &      71.09 &      0.394 &    0.84005 \\

       140 &  12942.183 &          1 &          6 &         12 &         20 &         29 &         72 &            &      71.50 &      0.362 &    0.83851 \\

       141 &  13136.029 &          1 &          6 &         13 &         20 &         29 &         72 &            &      71.87 &      0.375 &    0.84124 \\

       142 &  13331.394 &          1 &          6 &         13 &         20 &         30 &         72 &            &      72.22 &      0.384 &    0.84161 \\

       143 &  13528.267 &          1 &          6 &         13 &         20 &         30 &         73 &            &      72.56 &      0.381 &    0.84001 \\

       144 &  13726.522 &          1 &          7 &         13 &         20 &         30 &         73 &            &      72.94 &      0.379 &    0.84277 \\

       145 &  13926.409 &          1 &          7 &         13 &         21 &         30 &         73 &            &      73.32 &      0.392 &    0.84533 \\

       146 &  14127.660 &          1 &          7 &         13 &         21 &         30 &         74 &            &      73.68 &      0.369 &    0.84387 \\

       147 &  14330.350 &          1 &          7 &         14 &         21 &         30 &         74 &            &      74.03 &      0.391 &    0.84634 \\

       148 &  14534.569 &          1 &          7 &         14 &         21 &         31 &         74 &            &      74.39 &      0.387 &    0.84666 \\

       149 &  14740.224 &          1 &          7 &         14 &         21 &         31 &         75 &            &      74.74 &      0.387 &    0.84522 \\

       150 &  14947.576 &          1 &          7 &         14 &         21 &         32 &         75 &            &      75.04 &      0.376 &    0.84556 \\

       151 &  15156.337 &          1 &          7 &         14 &         22 &         31 &         76 &            &      75.45 &      0.380 &    0.84625 \\

       152 &  15366.486 &          1 &          7 &         14 &         22 &         32 &         76 &            &      75.81 &      0.378 &    0.84656 \\

       153 &  15578.084 &          2 &          8 &         14 &         22 &         31 &         76 &            &      76.12 &      0.390 &    0.85095 \\

       154 &  15791.167 &          2 &          8 &         14 &         22 &         32 &         76 &            &      76.48 &      0.385 &    0.85122 \\

       155 &  16005.627 &          2 &          8 &         14 &         22 &         32 &         77 &            &      76.76 &      0.360 &    0.84984 \\

       156 &  16221.679 &          2 &          8 &         15 &         22 &         32 &         77 &            &      77.10 &      0.368 &    0.85211 \\

       157 &  16439.236 &          2 &          8 &         15 &         22 &         32 &         78 &            &      77.47 &      0.378 &    0.85075 \\

       158 &  16658.206 &          3 &          8 &         15 &         22 &         32 &         78 &            &      77.79 &      0.384 &    0.85298 \\

       159 &  16878.650 &          3 &          9 &         15 &         22 &         32 &         78 &            &      78.16 &      0.372 &    0.85517 \\

       160 &  17100.602 &          3 &          9 &         15 &         22 &         33 &         78 &            &      78.47 &      0.387 &    0.85539 \\

       170 &            &            &            &            &            &            &            &            &      81.92 &      0.373 &            \\

       180 &            &            &            &            &            &            &            &            &      85.29 &      0.365 &            \\

       190 &            &            &            &            &            &            &            &            &      88.47 &      0.385 &            \\

       200 &  27192.287 &            &            &            &            &            &         92 &            &      91.81 &      0.382 &    0.87461 \\

       210 &            &            &            &            &            &            &            &            &      94.99 &      0.381 &            \\

       220 &            &            &            &            &            &            &            &            &      97.98 &      0.388 &            \\

       230 &            &            &            &            &            &            &            &            &     101.13 &      0.380 &            \\

       240 &            &            &            &            &            &            &            &            &     104.26 &      0.384 &            \\

       250 &            &            &            &            &            &            &            &            &     107.12 &      0.377 &            \\

       260 &            &            &            &            &            &            &            &            &     110.10 &      0.371 &            \\

       270 &            &            &            &            &            &            &            &            &     113.01 &      0.369 &            \\

       280 &            &            &            &            &            &            &            &            &     115.93 &      0.368 &            \\

       290 &            &            &            &            &            &            &            &            &     118.73 &      0.378 &            \\

       300 &  62862.152 &            &            &            &            &            &        122 &            &     121.56 &      0.377 &    0.90527 \\

       400 &  113558.53 &            &            &            &            &            &        149 &            &     148.22 &      0.363 &    0.92338 \\

       500 &  179375.07 &            &            &            &            &            &        172 &            &     172.77 &      0.364 &    0.93390 \\
       
\end{tabular}}

\normalsize
------------------------------------------------------------------------------}
\end{center}
\newpage
\begin{figure}
\begin{center}
\includegraphics[scale=0.4]{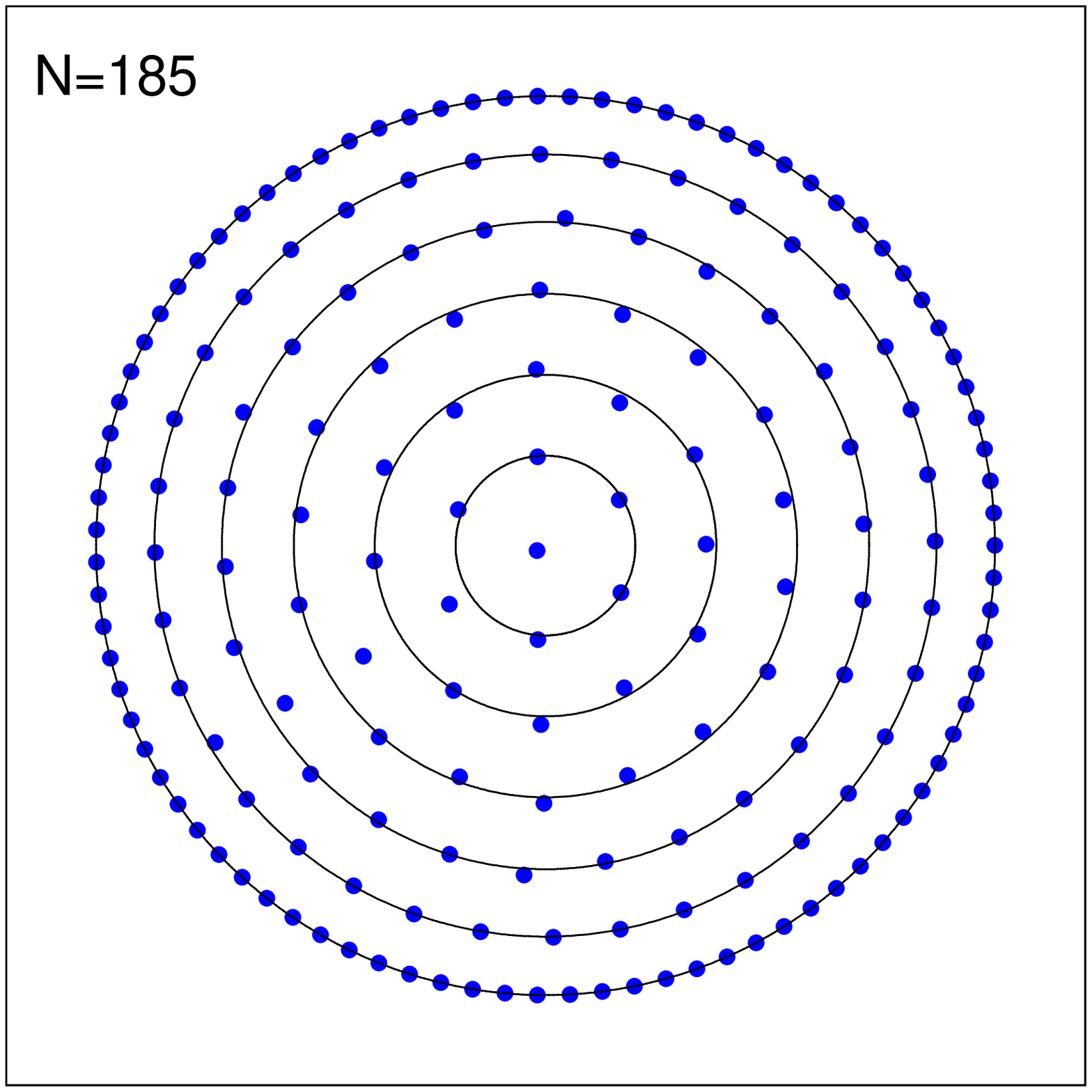}
\vspace{10pt}
\begin{minipage}[c]{0.8\linewidth}
\noindent \small Fig.3. The lowest energy configuration for $N=185$, the first $N$-value for which a simple concentric shell description appears to be inappropriate.  The configuration has $p=87$.
\end{minipage}
\end{center}
\end{figure}

\section{Total Energy}
Erko\c c and Oymak [16] have investigated the total energy, using the expression 
$$E(N)=\frac {\pi }{4}N^2+b_2N^{3/2}+b_3N \eqno(5)$$
The leading term is the exact expression for $E(N)$ as ${N \rightarrow \infty}$.  The remaining terms are the first two of an infinite series representing the correction for finite $N$. Erko\c c and Oymak fitted their data in the range ${2\leq N\leq 90}$, obtaining $b_2=-1.5599728$ and $b_3=0.9509338$.  Repeating the analysis with ${2\leq N\leq 160}$, we obtain $b_2=-1.5593651$ and $b_3=0.94290006$. These values predict $E(500)=179386.7591$ compared with the observed upper limit of $E(500)=179375.07$.  If a fit to a limited range of the data with ${100\leq N\leq 160}$ is performed, we obtain $b_2=-1.5611781$ and $b_3=0.9614771$.  These values predict $E(500)=179375.7771$, significantly closer to our best result for $N=500$.      

\section{Perimeter Particles}
If the shell structure is indeed lost as $N$ increases, it nevertheless seems likely that the perimeter particle count $p(N)$ will survive as a useful parameter and should be recognised as an integer sequence [17].  As support for such a sequence, an attempt has been made to express $p$ as a function of $N$ up to and beyond $N=160$.  This has been achieved as a by-product of an investigation into a faster method of identifying optimal configurations for higher values of $N$.

The increase in processing time required to obtain reliable solutions at higher $N$ suggests that some pre-preparation of the particle distributions may be advantageous.  Artificial distributions on their own are unlikely to deliver optimal solutions which are not, for the most part, made up of perfect circles and in any case the shell structure is unlikely to survive beyond some value of $N$.  Nevertheless, it was felt that such an approach could extend the range of reported results if suitable patterns of shells were used as inputs to the procedures described above.  If a sufficient number of different patterns are used as seeds, such that at least one is not separated from the optimal final state by a potential barrier, the need for the repeated annealing stages may be avoided. 

Table 1 and [10,11] all show 4-10-18-48 as the optimal configuration for $N=80$.  To generate such a sequence together with a number of nearby alternatives, the approach adopted here is to begin with $N$ charges on the disc in an equilibrium distribution of continuous charge.  We then regard the transition from such a distribution to a lowest-energy configuration of $N$ discrete charges as a process of condensation. To arrive at the correct final configuration it is necessary to establish the boundaries separating each unit charge on the continuous distribution. (An alternative method of establishing such boundaries, the Voronoi construction [18] has been applied to this problem by Bedanov and Peeters [13] who report results for hard-wall confinement with $N=50$ and $N=230$ only, obtaining a configuration of 1-5-13-31 for $N=50$.)  After integration of Eqn.1, the fraction of charge outside radius $r$ is given as $f(r)=(1-r^2)^{1/2}$ and between any two radii $r_{\rm a}$ and $r_{\rm b}$ with $r_{\rm b}\geq r_{\rm a}$ the fraction of the charge is $g=(1-{r_{\rm a}}^2)^{1/2}-(1-{r_{\rm b}}^2)^{1/2}$.  If such an area is divided into $n$ equal sectors such that each contains exactly the amount of charge to represent one particle, then $g/n=1/N$.  We begin at the perimeter, with all $N$ particles still available for allocation and $r_{\rm b}=1$, and ask how many particles $n$ should be assigned to the outermost shell.  Clearly at this stage the value of $n$ may fall in the range $1\leq n\leq N$.  If $n=N$ is chosen then $r_{\rm a}=0$, but for all other values we obtain $0<r_{\rm a}<1$ (Fig.4).  For all possible values of $n$ we calculate the length of the boundary of the cell representing one particle and accept the value of $n$ giving the smallest boundary.  This defines $p=n$.  The corresponding value of $r_{\rm a}$ is then taken as the outer radius contributing charge to the next shell, the remaining number of particles is adjusted to $N-n$ and the process is repeated until no particles remain.
  
\begin{figure}
\begin{center}
\includegraphics[scale=0.35]{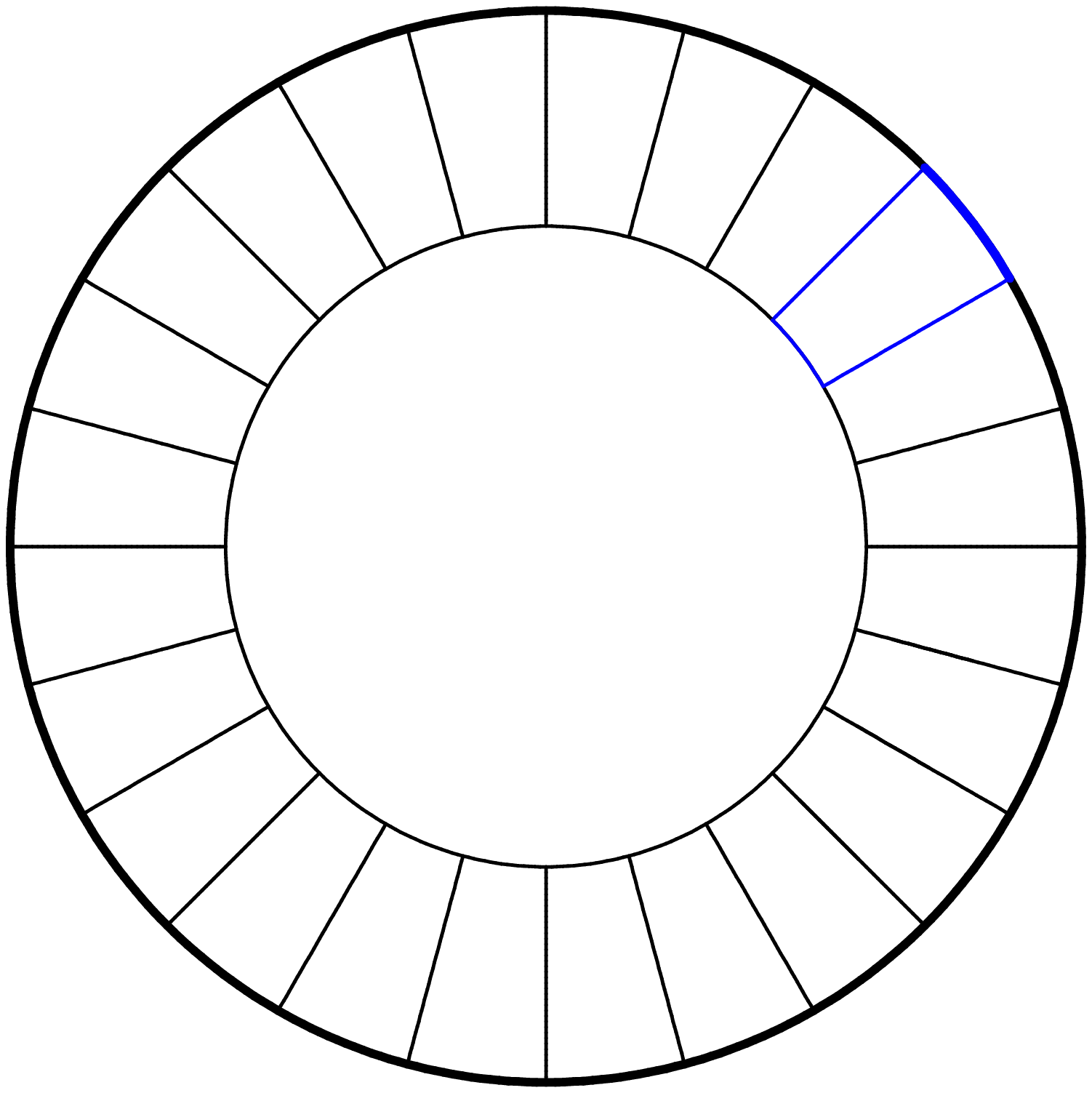}
\vspace{10pt}
\begin{minipage}[c]{0.80\linewidth}
\noindent \small {Fig.4. Working inwards from the perimeter, various numbers of particles (initially ${1\leq n\leq N}$) are allocated to each shell such that for each choice of $n$, each cell contains a single charge, using the distribution given by Eqn.1.  The number adopted is the one producing the lowest cell perimeter (shown in blue), taking into account the weighting factor which multiplies the radial component. The procedure is then repeated for the next inner shell until no particles remain.}
\end{minipage}
\end{center}
\end{figure}

Applying the above procedure for $N=80$, we obtain 2-8-14-22-34.  For $N=160$ the configuration (Table 1) is 3-9-15-22-33-78, and the procedure yields 1-7-13-20-28-37-54.  But it was noted that by making a simple modification to the calculation of the boundary, initially for the first (perimeter) shell only, much closer agreement could be obtained. A weighting factor $w$ was introduced for the radial part of the boundary, so that its effective circumference $s$ was taken as
$$s=2w(r_{\rm b}-r_{\rm a})+2\pi (r_{\rm b}+r_{\rm a})/n\eqno(6)$$
and the minimum value of $s$ is used to select $r_{\rm a}$.  With $w_1=0.3$ for the first shell and $w_2,w_3...=1.0$ for all subsequent shells, the new configuration obtained for $N=80$ is 4-10-17-49, and for $N=160$, 3-9-15-23-31-79.
   
A fuller investigation was then undertaken to establish the values of the weights that generate the \textit{exact} optimal configuration for each $N$-value in the range ${20\leq N\leq 160}$.  At a particular value of $N$, $w_1$ for example can take a range of values and still deliver the correct $p$, so each $N$ produces a maximum and a minimum value for each weight. The results for the perimeter weight $w_1$ are shown in Fig.5, which also includes the results for selected values with $N>160$ obtained from $N$ and $p$ only. It is found that as $N$ increases, the correct perimeter count $p$ may be obtained with a value of $w_1$ close to 0.30. To generate a collection of candidate configurations for further processing by the relaxation method, it is only necessary to repeat the condensation routine with weights $w_1$,$w_2$... each varying in a narrow range about its adopted central value.  Further results beyond $N=160$ using this method will be presented in future work.  

\begin{figure}
\begin{center}
\includegraphics[scale=0.4]{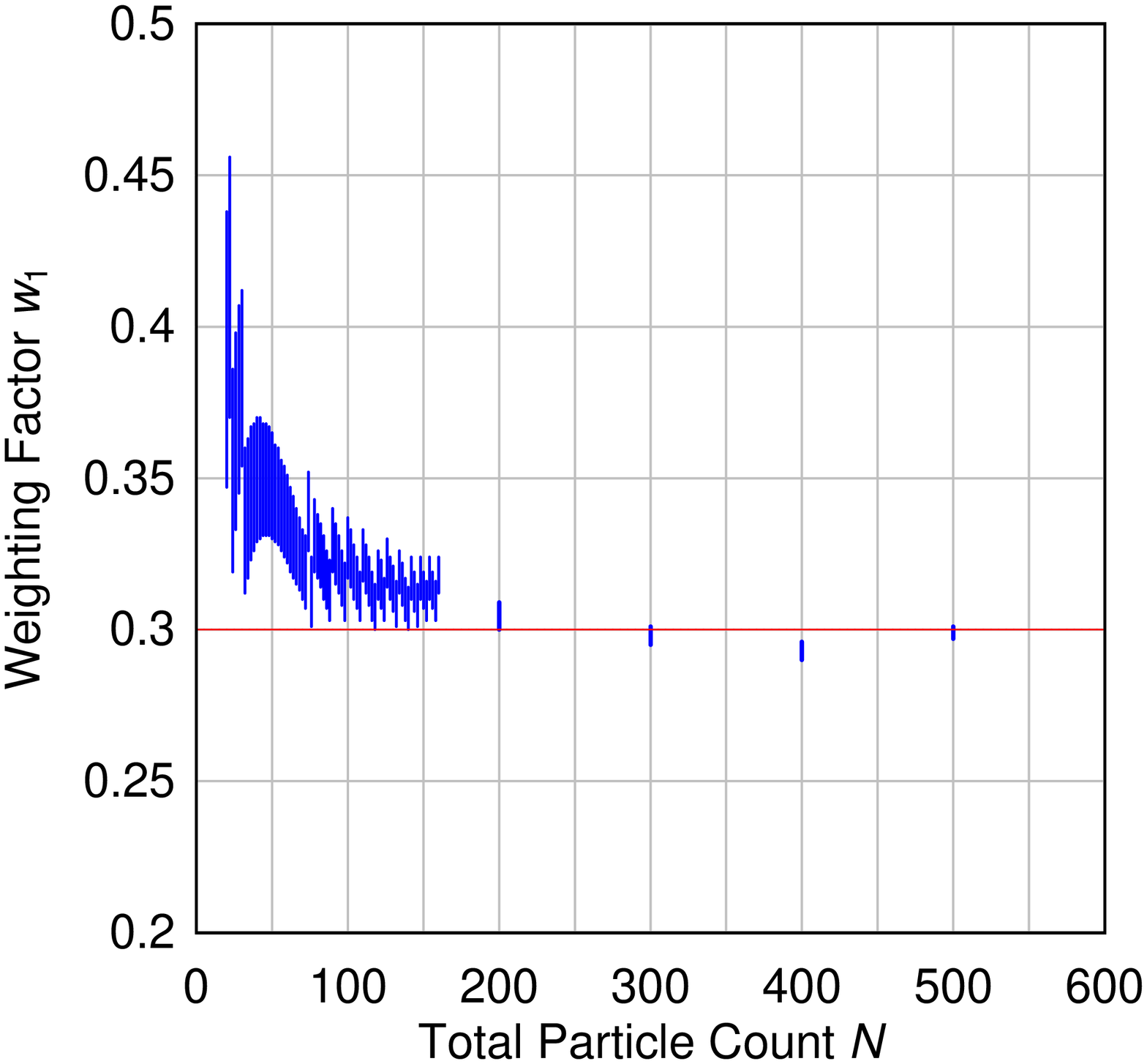}
\vspace{10pt}
\begin{minipage}{0.80\linewidth}
\noindent \small Fig.5. The range of values of $w_1$ that generate the correct configuration for ${20\leq N\leq 160}$ and selected values up to $N=500$.  The results for $N>160$ are obtained without knowledge of the full configuration and use $N$ and $p\;$ to measure $w_1$ only.
\end{minipage}
\end{center}
\end{figure}
 
\noindent The eventual similarity of the results for $w_1$ suggests that the consequences of a fixed value of this weight should be investigated. We assume that there is some radius $r$ such that the fraction of charge outside $r$ accounts for the perimeter particles, and write $p=N(1-r^2)^{1/2}=N({1-r)}^{1/2}{(1+r)}^{1/2}$.  Substituting this version of $p$ into Eqn.6 and setting $n=p$, $r_{\rm a}=r$ and $r_{\rm b}=1$, we obtain
$$s=2w_1(1-r)+{\frac {2\pi }{N}}{\left({\frac {1+r}{1-r}}\right)^{1/2}}$$
Hence, to select the minimum boundary length

$${\frac {ds}{dr}}=-2w_1+{\frac{2\pi}{N}}{{(1-r)}^{-3/2}{(1+r)}}^{-1/2}$$
$$=-2w_1+{\frac{2\pi}{p(1-r)}}$$
$$=0\hspace{10pt}{\rm for}\hspace{10pt} p(1-r)=\frac{\pi}{w_1}\eqno(7)$$

\vspace{10pt}
If we take $w_1=0.3$ and identify $r$ as the radius of the outermost inner shell $r_2$, minimising the boundary leads to the relation $p(1-r_2)=10.47$.   Fig. 6 shows the variation of $p(1-r_2)$ with $N$ for all the results shown in Table 1. The mean value in the range ${100\leq N\leq 160}$ is 11.461, and the higher values of $N$ remain close to this value.  The quantity $p(1-r_2)$ appears to tend to a constant value, though we note that $r$ refers to the inner boundary of the region producing the perimeter particles, and hence the outer boundary contributing the second shell.  This radius is not necessarily the radial position of the second shell particles $r_2$, so the values of the constants need not agree.  

\begin{figure}
\begin{center}
\includegraphics[scale=0.4]{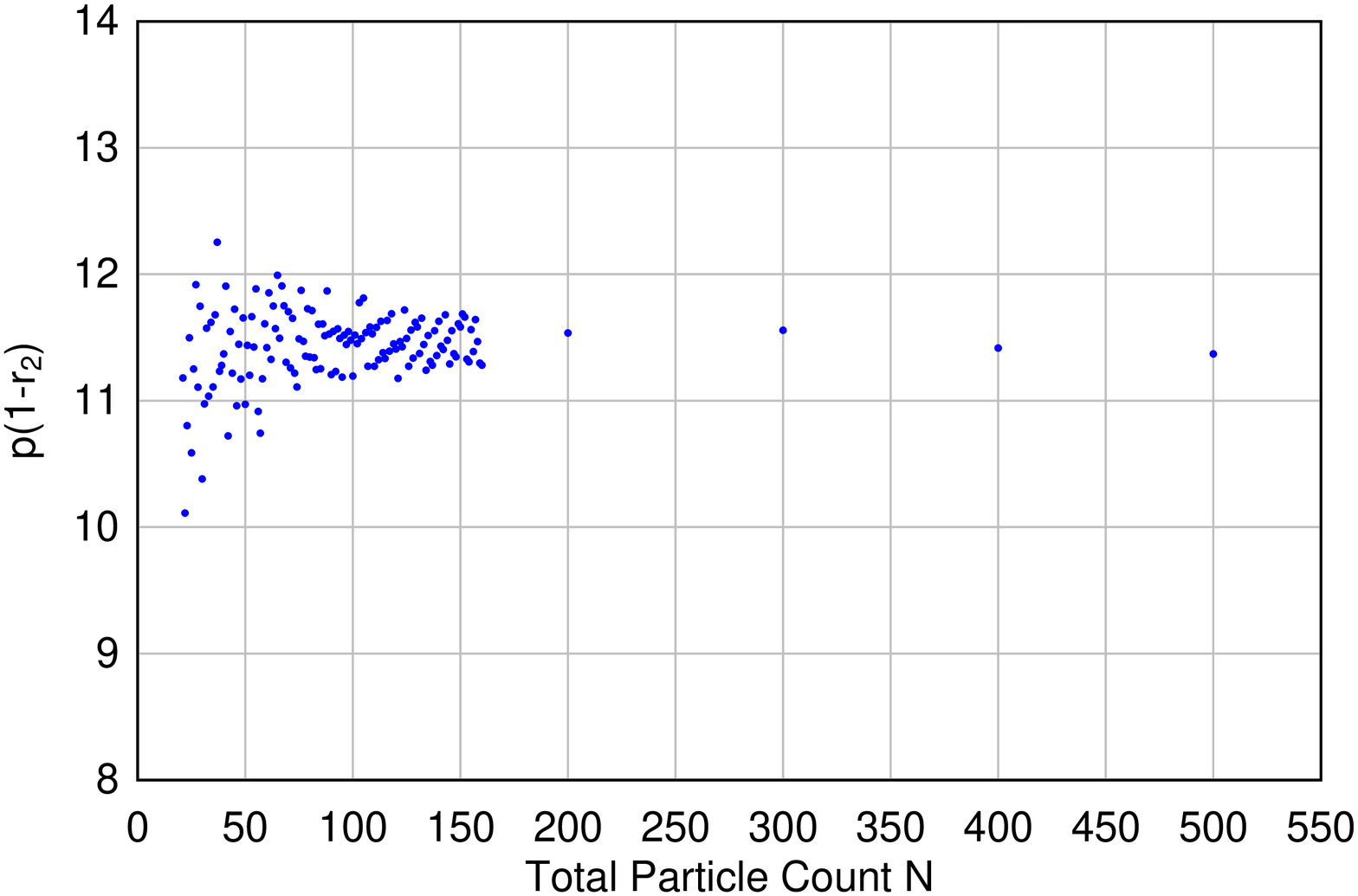}
\vspace{10pt}
\begin{minipage}[c]{0.80\linewidth}
\noindent \small Fig.6. The relationship between $p$ and $r_{\rm 2}$ for ${21\leq N\leq 160}$ and for selected values of $N$ up to $N=500$.
\end{minipage}
\end{center}
\end{figure}    

Finally, substituting for the remaining $r$ in Eqn. 7 produces
$$p{\left(1-(1-{p^2}/{N^2})^{1/2}\right)=\frac{\pi}{w_1}}$$
  
\noindent For large $N$ such that $p/N\ll 1$ the square root term may be replaced by its three leading terms, resulting in a quintic equation in $p$.  The two leading terms of the solution are
$$p={\left(\frac{2\pi}{w_1}\right)}^{1/3}{N}^{2/3}-{\frac{1}{6}}{\left(\frac{\pi}{w_1}\right)}\eqno(8)$$

\vspace{10pt} 
\noindent If we again take $w_1=0.3$, Eqn.8 becomes $p=2.76N^{2/3}-1.75$.  Using the results for all $N$ listed in Table 1, Fig.7 shows the dependence of $p_{\rm min}$ upon $N^{\rm 2/3}$, yielding a best fit of $p=2.7922N^{2/3}-3.723\pm 0.021$.  Once again, the results are not strictly comparable - both use the same $p$ and $N$ data, but Eqn.8 relies upon our assumption that $p/N\ll 1$.  Nevertheless, the simple method of charge partitioning using only the single parameter $w_1$ as an input produces a relationship between $p$ and $N$ which appears realistic.  The estimated error on the fitted constant term was then used with the coordinates of the centroid of our data ($N^{\rm 2/3}=21.7988$, $p_{\rm min}=57.1445$) to predict values of $p_{\rm min}$ beyond $N=500$.  We obtain $p_{\rm min}=(3.723\pm 0.021)s+t$ where $s=0.04587N^{\rm 2/3}-1$ and $t=2.6215 N^{\rm 2/3}$. Although such extrapolations must clearly be treated with caution, we have used this to estimate the range of $p_{\rm min}$ for $N=1000$, obtaining $p_{\rm min}=275.50\pm 0.08$.  Efforts with the spectrum program have so far yielded a best energy of $E=736985.16$, obtained at $p=276$.  This $p$-value must be regarded as preliminary and the energy as an upper limit, but the relation between $p$ and $N$ does appear to be quite robust.  If $p(1000)=276$, this requires (Fig.5) $w_1$ in the range 0.292-0.294.
 
\begin{figure}
\begin{center}
\includegraphics[scale=0.4]{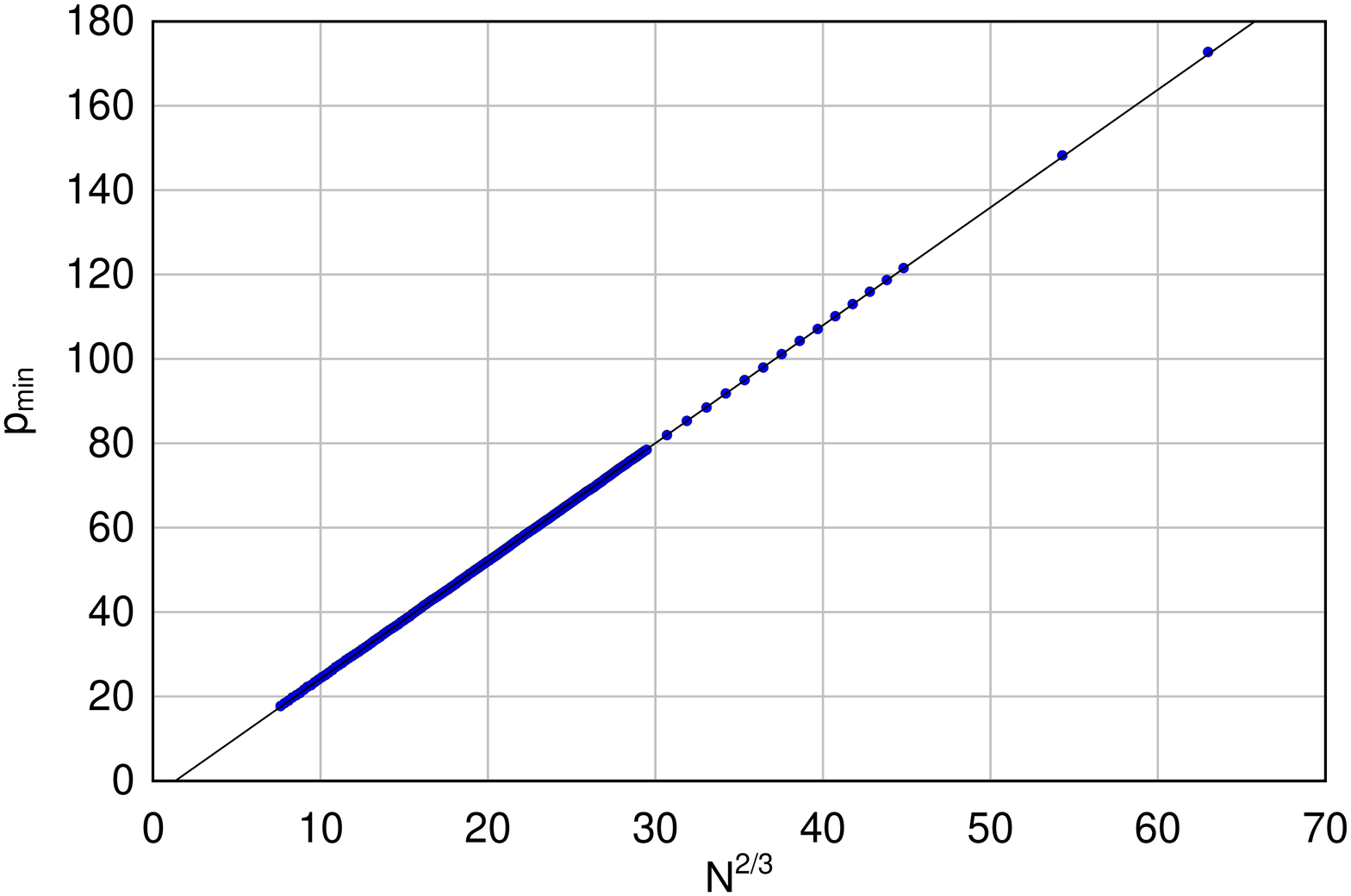}
\vspace{10pt}
\begin{minipage}[c]{0.80\linewidth}
\noindent \small Fig.7. The variation of $p_{\rm min}$ with $N^{\rm 2/3}$.  The typical standard deviation on a measurement of $p_{\rm min}$ is 0.04.
\end{minipage}
\end{center}
\end{figure}  
Atiyah and Sutcliffe [19] have designated the $N$-values at which the latest particle is added to the interior of the circle rather than to the perimeter as 'jumping points'.  The sequence of such points begins $N=12,17,19,22...$ and in our notation these are the points of unchanging $p$.  This suggested an alternative method of fitting Eqn. 8 by searching for $c_1$ and $c_2$ such that the nearest integer to ${c_1}N^{2/3}-c_2$ gave the correct value of $p$ in the highest possible number of cases.  It was found that in the range ${21\leq N\leq 160}$ all but eight values of $p$ could be correctly generated with $c_1=2.7777$ and $c_2=3.3203$.  If a 100\%  success rate had been achieved, the jumping points would have been directly calculable from the expression for the $p$-values.        
\section{Conclusions}
We have obtained solutions for the optimal distributions of $N$ charged particles on a unit disc, extending the known configurations to $N=160$.  Although it appears that well-behaved patterns of concentric shells may become less prevalent after $N\sim 185$, it is suggested that the perimeter count $p(N)$ should survive as a useful parameter, and tentative values for $p_{\rm min}$ have been obtained for further selected values of $N$ up to $500$.  Over the range studied $p$ varies according to $p=2.7922{N^{\rm 2/3}}-3.723$.  If this expression is used to determine $p(1000)$ the result appears to be supported by a preliminary calculation. A simple procedure for assigning particles to concentric shells appears to support the observed relationships between $p$ and $N$ and between $p$ and $r_2$.   

\section*{Acknowledgement}
I would like to thank Jennifer Scherr for valuable assistance in tracing several useful references.                                                                        

\section*{References}
[1] M.G.CALKIN, D.KIANG and D.A.TINDALL, Minimum energy charge configurations, {\sl Am.J.Phys.} {\bf 55} (1987) 157-158.

\noindent[2] J.J.THOMSON, On the structure of the atom {\sl Phil. Mag.} {\bf 7} (1904) 237.

\noindent[3] J.DE LUCA, S.B.RODRIGUES and Y.LEVIN, Electromagnetic instability of the Thomson problem, {\sl Europhysics Letters} {\bf 71} (2005) 84-90.

\noindent[4] E.L.ALTSCHULER, A.PEREZ-GARRIDO, Global minimum for Thomson's problem of charges on a sphere, {\sl Phys. Rev. E} {\bf 71} (2005) 047703.

\noindent[5] A.A.BEREZIN, An unexpected result in classical electrostatics, {\sl Nature} {\bf 315} (1985) 104.

\noindent[6] R.NITYANADA, Electrostatics vindicated..classically, {\sl Nature} {\bf 316} (1985) 301.

\noindent[7] M.REES, The distribution of charges in classical electrostatics, {\sl Nature} {\bf 317} (1985) 208.

\noindent[8] R.FRIEDBERG, The electrostatics and magnetostatics of a conducting disc, {\sl Am.J.Phys},{\ }{\bf 61} (1993) 1084-1096.

\noindent[9] J.D.JACKSON, {\sl Classical Electrodynamics} (Wiley, New York, 1962,5,7), Section 3.12.

\noindent [10] K.J.NURMELA, Minimum-energy point charge configurations on a circular disc, {\sl J.Phys.A.} {\bf 31} (1998) 1035-1047. 
 
\noindent [11] H.OYMAK and \c S.ERKO{\c C}, Distribution of point charges on a thin conducting disc, {\sl Int.J.Mod.Phys.C},{\ }{\bf 5} (2000) 891-900.

\noindent [12] M.SAINT JEAN, C.EVEN, C.GUTHMANN, Macroscopic 2D Wigner islands, {\sl Europhysics Letters} {\bf 55} (2001) 45-51.

\noindent [13] V.M.BEDANOV and F.M.PEETERS, Ordering and phase transitions of charged particles in a classical finite two-dimensional system, {\sl Phys. Rev. B} {\bf 49} (1994) 2667-2676.

\noindent [14] L.T.WILLE and J.VENNIK, Electrostatic energy minimisation by simulated annealing, {\sl J.Phys.A.} {\bf 18} (1985) L1113-1117 and corrigendum {\bf 19} (1986) 1983.   

\noindent[15] L.BONSALL and A.A.MARADUDIN, Some static and dynamical properties of a 2-D Wigner crystal, {\sl Phys.Rev.B} {\ }{\bf 15} (1977) 1959-1973.

\noindent [16] \c S.ERKO{\c C} and H.OYMAK, Energetics and stability of discrete charge distribution on a conducting disc, {\sl Physics Letters A},{\bf 290} (2001) 28-34. 

\noindent[17] N.J.A.SLOANE, The online encyclop\ae dia of integer sequences, found at URL:
  
http://www.research.att.com/$\sim $njas/sequences/

\noindent [18] A.OKABE, B.BOOTS, K.SUGIHARA and S.N.CHIU, Spatial Tessalations: Concepts and Applications of Voronoi Diagrams, Wiley (2000).

\noindent [19] M.ATIYAH and P.SUTCLIFFE, The geometry of point particles, {\sl Proc. R. Soc. Lond. A} {\bf 458} (2002) 1089-1115. 
 
\end{document}